\documentclass{article}

\usepackage{PRIMEarxiv}

\usepackage[utf8]{inputenc} % allow utf-8 input
\usepackage[T1]{fontenc}    % use 8-bit T1 fonts
\usepackage{hyperref}       % hyperlinks
\usepackage{url}            % simple URL typesetting
\usepackage{booktabs}       % professional-quality tables
\usepackage{amsfonts}       % blackboard math symbols
\usepackage{nicefrac}       % compact symbols for 1/2, etc.
\usepackage{microtype}      % microtypography
\usepackage{lipsum}
\usepackage{fancyhdr}       % header
\usepackage{graphicx}       % graphics
\graphicspath{{media/}}     % organize your images and other figures under media/ folder
\usepackage{amsmath}
\usepackage{subfig}
\usepackage{makecell}

\title{Don't Forget To Look Up
}

\author{
  Philip Lubin and Alexander N. Cohen\\
  Department of Physics \\
  University of California -- Santa Barbara \\
  Santa Barbara, CA 93106\\
  \texttt{lubin@ucsb.edu} \\
}

\begin{document}
\maketitle

\begin{abstract}
We discuss a hypothetical existential threat from a 10 km diameter bolide discovered 6 months prior to impact with one case being a comet and the other being an asteroid. We show that an extension of our work on bolide fragmentation using an array of penetrators but modified with small nuclear explosive devices (NED) in the penetrators, combined with soon-to-be-realized heavy lift launch assets with positive $C_3$ such as NASA SLS or SpaceX Starship (with in-orbit refueling) is sufficient to mitigate this existential threat. A threat of this magnitude hitting the Earth at a closing speed of 40 km/s for the comet and 20 km/s for the asteroid would have an impact energy of roughly 65 Teratons TNT, or about ten thousand times larger than the current combined nuclear arsenal of the entire world. This is similar in energy to the KT extinction event that killed the dinosaurs some 66 million years ago. Such an event, if not mitigated, would be an existential threat to humanity. We show that mitigation is conceivable using existing technology, even with the short time scale of 6 months warning, but that the efficient coupling of the NED energy is critical.
\end{abstract}

% keywords can be removed

\section{Introduction}

\subsection{Asteroid and Comet Impact Threat}

\begin{figure}
    \centering
    \includegraphics[width=0.6\textwidth]{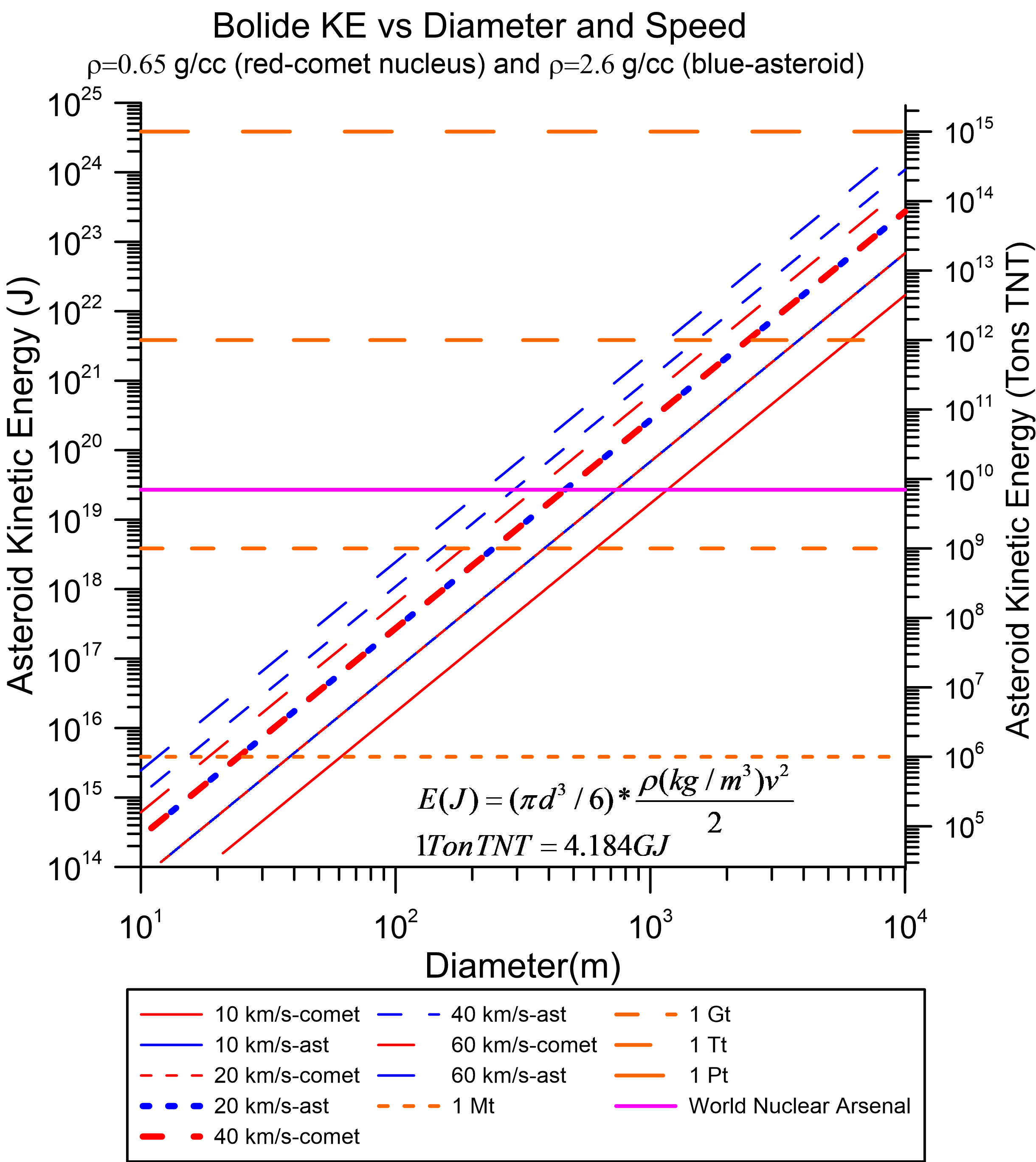}
    \caption{Exo-atmospheric kinetic energy vs. diameter and atmospheric entry speed for both typical comet and asteroid densities (0.65 and 2.6 g/cm$^3$, respectively). Relevant energy scales and total human nuclear arsenal shown for comparison. For diameters greater than 500 m, the bolide energy exceeds the world’s nuclear arsenal. Humanity is good at building weapons, but nature is far better at it.}
    \label{fig:KEvsdiameter}
\end{figure}

Asteroid and comet impacts pose a continual threat to life on Earth. For example, on February 15, 2013, a 20 m diameter asteroid airburst over Chelyabinsk, Russia with a total energy of about 0.5 Mt, or roughly that of a modern ICBM thermonuclear warhead \cite{popova_chelyabinsk_2013}. About 50\% of its energy went into the atmospheric blast wave that injured approximately 1,600 people. The same day, a 50 m diameter asteroid (2012 DA14) passed within the geosynchronous satellite belt. Had it impacted, it would have had a yield of about 10 Mt. Such an impact over a major city could have killed and injured millions. Asteroids the size of 2012 DA14 ($\sim50$ m diameter) are expected to strike Earth approximately every 650 years, while objects at least the size of the Chelyabinsk asteroid ($\sim20$ m diameter) are expected to strike Earth approximately every 50--100 years \cite{glasstone_effects_1977}. Another recent example is the Tunguska 1908 event that could have caused large scale loss of life, but did not due to the remote area in Russia it airburst over. This event is estimated to have been a roughly 65 m diameter asteroid (or possibly an atmospheric grazing comet) and resulted in an air blast with total energy yield of between 3 and 30 Mt. The resulting blast wave flattened more than 2000 km$^2$ of forest, as seen from the 1927 Soviet Academy of Sciences expedition.  Larger objects pose an even more severe threat. For example, the total kinetic energy associated with an impact of a 100 m asteroid is equivalent to approximately 100 Mt, and that of the well-known $\sim350$ m threat, Apophis, would have an impact yield of approximately 3--4 Gt \cite{brown_flux_2002}, or about 1/2 of the Earth’s total nuclear arsenal, while Bennu at 490 m could have a yield greater than the entire Earth’s nuclear arsenal. This is summarized in Figure \ref{fig:KEvsdiameter}, wherein we show the bolide kinetic energy as a function of its diameter. For reference, Apophis will next visit Earth on Friday April 13, 2029 and come within the geosynchronous belt. 

Effective mitigation strategies are imperative to ensure humanity’s continuity and future advancement. Every day, approximately 100 tons of small debris impact the Earth and the effect is virtually undetectable. Extremely large asteroids ($>1$ km diameter) and comets pose unique threats. Existential threats to humanity are very low recurrence, but are known to have happened multiple times in the past including the last mass extinction approximately 65 Myr ago when an estimated 10 km diameter asteroid triggered large scale extinction of life (KT boundary), including the destruction of dinosaurs \cite{alvarez_extraterrestrial_1980}. The equivalent yield of this event is estimated to have been approximately 100 Tera-tons TNT, or about 15,000 times larger than the current nuclear arsenal.  In our recent work on terminal planetary defense, referred to collectively as ``PI,'' we focused on non-existential but still extremely dangerous threats. In this paper we focus on a hypothetical threat from a large asteroid or comet with very short warning time that would have truly existential consequences for life on Earth, particularly for humanity. Such an event is often referred to as a ``planet killer.'' We focus on a hypothetical 10 km diameter threat that is detected 6 months prior to impact.  We explore multiple cases, but focus on both a comet with nucleus density of 0.65 g/cm$^3$ at a closing speed speed of 40 km/s (0.69 AU/month, or 4.1 AU in 6 months) and a comet with density 2.6 g/cm$^3$  with a closing speed of 20 km/s (0.35 AU/month, or 2.1 AU in 6 months). Both have the same kinetic energy of about 65 Teratons TNT. Both would be considered existential threats, or planet killers. We ignore the detailed orbital dynamics as this is case specific. The question we address is ``could we mitigate the threat and save humanity?''

\begin{figure}
    \centering
    \includegraphics[width=0.6\textwidth]{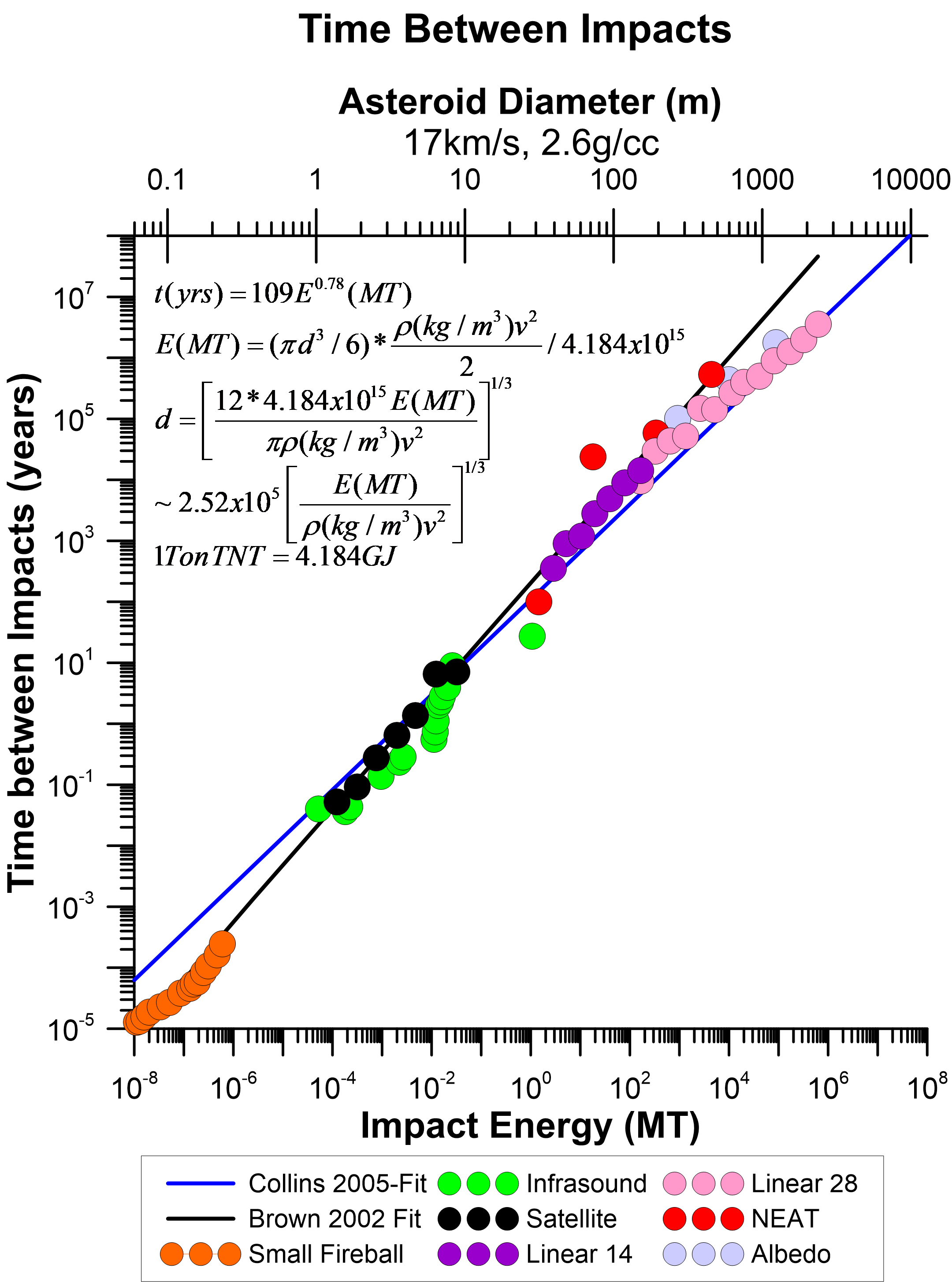}
    \caption{Approximate power law relationship between impact recurrence times and exo-atmospheric impact kinetic energy. Based on fitting known and extrapolated impacts from Taylor and Zeldovich \cite{taylor_formation_1950-1,taylor_formation_1950,zeldovich_theory_1950}.}
    \label{fig:timebetweenimpacts}
\end{figure}

While a ``planet killer'' event is quite rare, of order once per 100 million years, as shown in Figure \ref{fig:timebetweenimpacts}, such events will happen again in the future \cite{taylor_formation_1950-1,taylor_formation_1950,zeldovich_theory_1950}. As we discuss in detail in our recent paper, a robust planetary defense system should not focus only on rare existential threats, but should focus on the much larger number of smaller but deadly threats, particularly those from about 20--500 m in diameter. In the process of doing so, we can use the same basic system for much larger threats if needed. We show that an approach using an array of hypervelocity penetrators that are either inert or filled with conventional explosives, is extremely effective in enabling both rapid short term response to small diameter threats, as well as to larger ones (up to 1 km diameter). In the case of our 10 km ``planet killer,'' we show that using the same basic system also works when the penetrators are equipped with nuclear explosive devices (NED’s). In this sense we show that humanity has crossed a technological threshold to prevent us from ``going the way of the dinosaurs.''

\section{Existential Threats}

\begin{figure}
    \centering
    \includegraphics[width=0.475\textwidth]{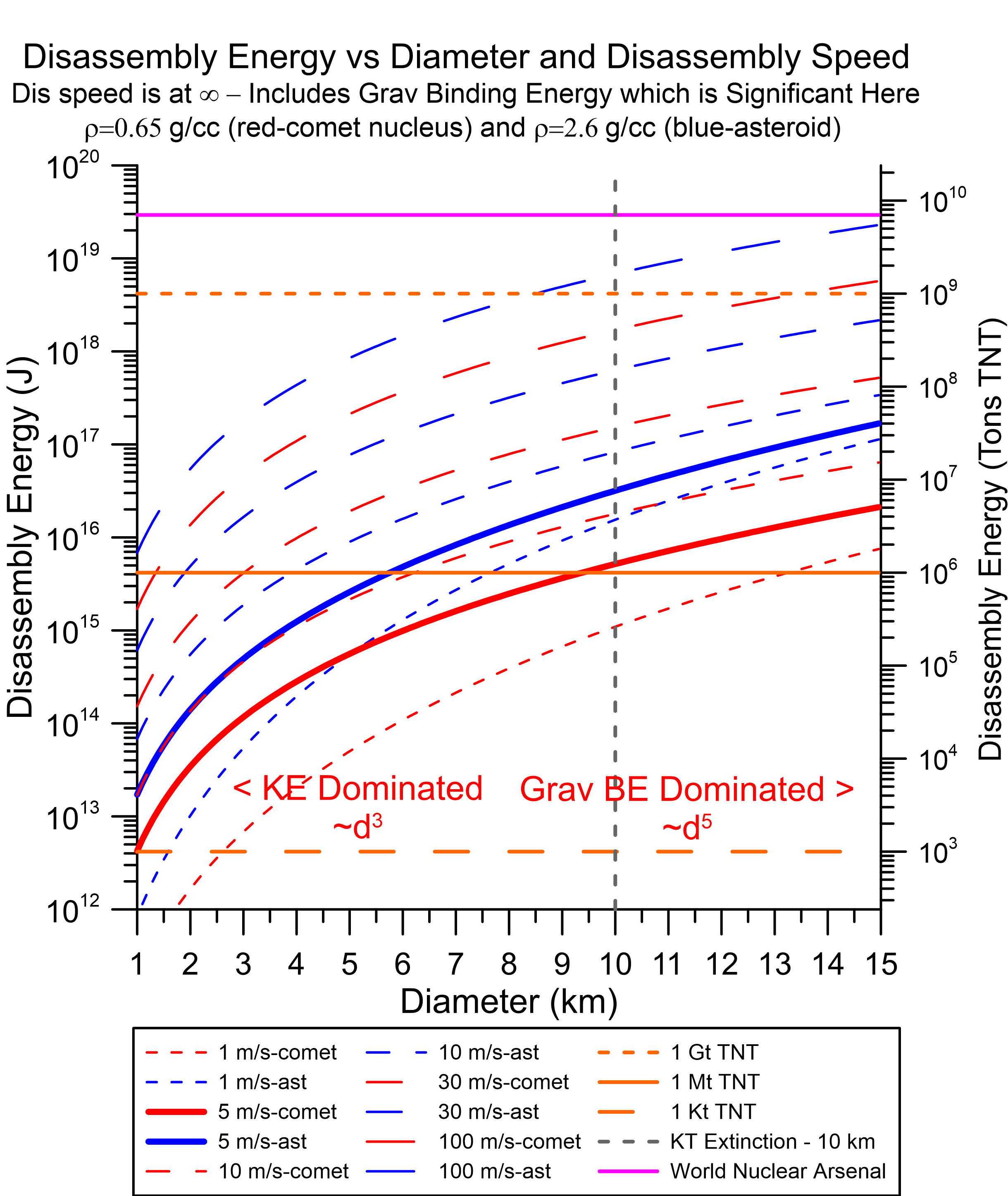}
    \includegraphics[width=0.475\textwidth]{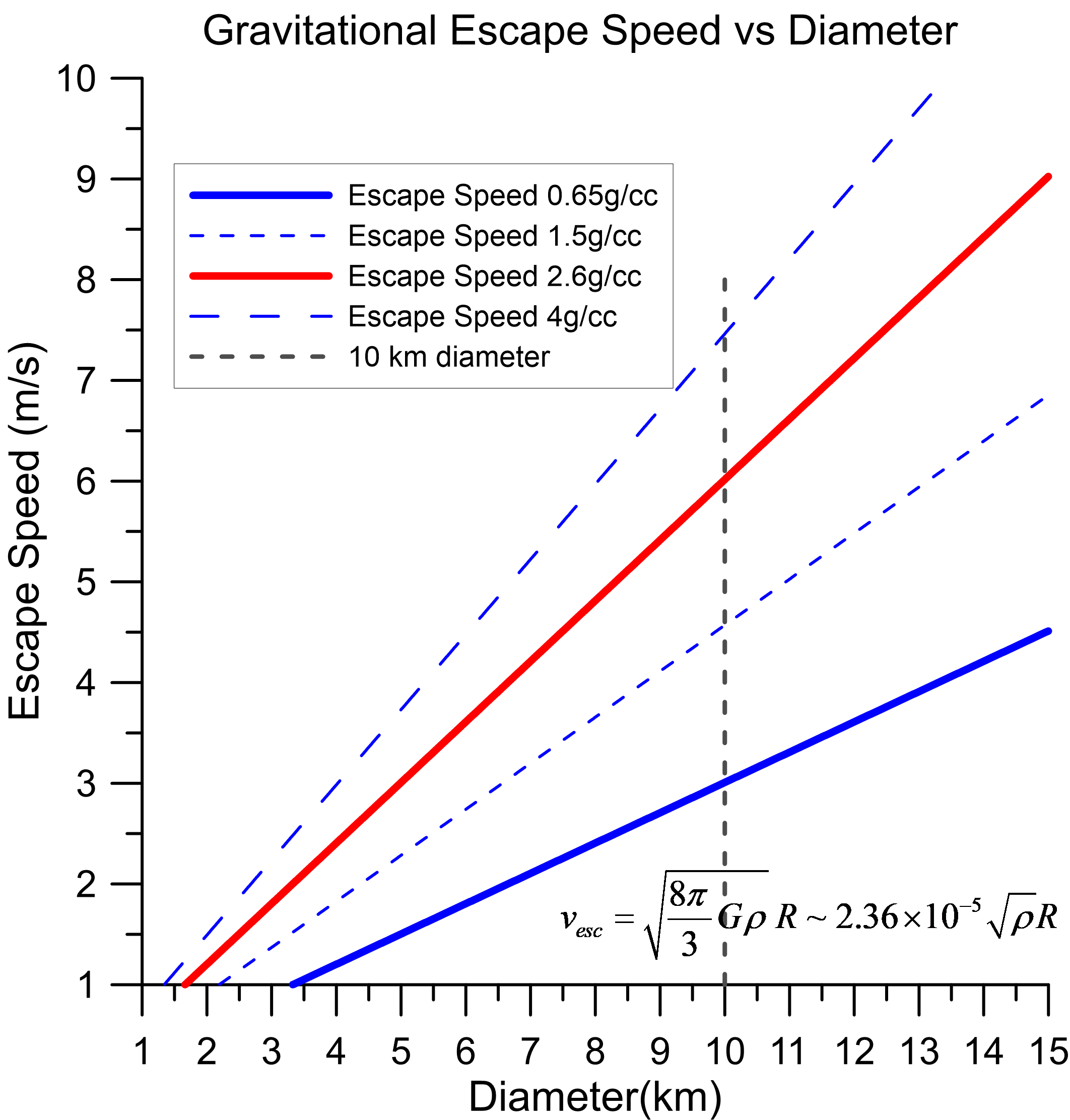}
    \caption{(Left) Energy required to disassemble a large asteroid or comet at a given fragment speed relative to the center of mass, for fragment speeds from 1 to 100 m/s at infinity and parent comet and asteroid diameters from 1 to 15 km. Note that unlike the case of small bolides, the case of large bolides has very significant gravitational binding energy for low disassembly speeds (at infinity). Above 40 km diameter, there is not sufficient explosive energy in the world’s current arsenal to even gravitationally de-bind the bolide for asteroids. (Right) Escape speed vs. diameter vs. density. Solid red and blue lines show the asteroid and comet cases, respectively.}
    \label{fig:disassemblyenergy}
\end{figure}

The basis for the PI method (standing for ``Pulverize It'') is to use an array of penetrators that injects detonation energy within the bolide to fragment, gravitationally de-bind, and add sufficient energy to spread the resulting fragment cloud \cite{lubin_asteroid_2022,lubin_pi_2022,lubin_planetary_2021}. For smaller threats up to about 1 km diameter, all the fragments can hit if intercept time is short as the Earth’s atmosphere acts as ``body armor'' and the resulting blast waves are de-correlated, which mitigates the threat \cite{lubin_pi_2022}. For threats larger than 1 km, the system is the same, but the threat is fragmented early enough so that the fragment cloud spreads to become larger than the Earth and thus the fragments miss the Earth. Any residual fragments that are threatening can be dealt with. The general process is to completely fragment the bolide into small enough fragments to avoid catastrophe. The minimum injection energy required is given by
\begin{equation}
    E=\textrm{KE}+G_\textrm{BE}=\frac{\pi d^3}{6}\frac{\rho v^2}{2}+\frac{1}{30}G\pi^2\rho^2 d^5,
\end{equation}
where $G_\textrm{BE}$ is the gravitational binding energy of the original bolide, and $d$ [m], $\rho$ [kg/m$^3$], and $v$ [m/s] are the original bolide diameter, density, and velocity, respectively. The speed $v$ is at infinity, or far from the original bolide. The gravitational escape speed from the surface of the bolide is given by
\begin{equation}
    v_\textrm{esc}=\sqrt{\frac{8\pi}{3}G\rho}\mathop{R}\sim 2.36\times10^{-5}\sqrt{\rho}\mathop{R}.
\end{equation}
To achieve a speed $v_\infty$ at infinity, we need to have an initial speed at disassembly for the outer layer of material of
\begin{equation}
    v_\textrm{int}=\sqrt{v_\textrm{esc}^2+v_\infty^2}.
\end{equation}
As we remove the outer layers of the bolide and move towards the center, the escape speed decreases to zero. The key is to inject energy in such a way that it couples relatively efficiently into the resulting outward kinetic energy of the fragments. The typical density for a comet nucleus is estimated to be $0.6\pm0.2$ g/cm$^3$ \cite{weissman_structure_2008}, while a typical stony asteroid has a density of about $2.6\pm1$ g/cm$^3$. We analyze both cases below allowing a comparison of a comet and asteroid threat of the same size but at different speeds relevant for each case. The PI system has multiple modes of operation so that even if some fragments will hit the Earth these additional modes can mitigate these smaller fragment as well, if required. This is discussed in detail in our recent paper. As shown in Figure \ref{fig:disassemblyenergy}, the energy required for complete disruption is dominated by the gravitational binding energy for bolides larger than a few km in diameter and for fragment speeds far from the original bolide of 1 m/s. A fragment speed of $\sim1$ m/s is sufficient to miss the Earth assuming an intercept of greater than roughly 2.5 months (75 days). In the worst case of an Earth-grazing bolide, a fragment speed of 2 m/s and 75 days suffices. This is true for any bolide size, though for smaller bolides we can ``afford'' to use the Earth’s atmosphere as an effective shield if the fragments are less than 15 m in diameter. There is a tradeoff in intercept time prior to impact and the speed required of the fragments. This becomes an energy input trade-off. As another example (our case), an intercept at one month prior to impact would require a fragment dispersal speed of 5 m/s to have the fragment cloud radius be equal to the Earth’s diameter.

\subsection{Can the Earth Simply Absorb the Bolide Energy?}

There are three main effects from a bolide impact, even if fragmented to a size scale less than 15 m in diameter, and preferably smaller:
\begin{itemize}
    \item Acoustic signature from atmospheric blast waves from the fragments;
    \item Optical signature pulse from the friction in the atmosphere on the fragments;
    \item Dust production in the atmosphere from the fragment burn-up -- nuclear winter effects;
    \item Atmospheric temperature rise due to the bolide energy injection ultimately being dissipated as heat.
\end{itemize}

As discussed extensively in our PI paper, bolide threats below about 1 km can be mitigated solely ``within the Earth’s atmosphere,'' while larger threats have too much energy to be safely absorbed. However, when it comes to truly existential threats with diameters greater than about 5 km, the amount of energy released in the acoustic and optical signatures and even the dust production becomes so large as to overwhelm our atmospheric protective shield. For large diameter existential threats, it is necessary to have the great majority of the threat miss the Earth completely. This is discussed in detail in our PI paper.

\subsection{Comparison Between Comet Nuclei and Asteroids}

We compare the key differences in gravitational binding and disruption energy for the two cases of the 10 km diameter comet nucleus of density 0.65 g/cm$^3$ with speed 40 km/s and the same sized asteroid with density 2.6 g/cm$^3$ and speed 20 km/s. Both cases are assumed to have 5 m/s fragment speed at infinity to allow fragments to miss the Earth with one month prior to impact intercept. Both have a total kinetic energy of about 65 Teratons and cause an average atmospheric temperature rise of about 55 C if the total KE energy of an impact were completely absorbed by the Earth’s atmosphere. The average atmospheric temperature rise is only for a point of discussion to indicate that mitigation via fragmentation and absorption in the atmosphere as is done in our PI program for non-existential threats is NOT an option for extremely large threats such our 10 km threat in this paper. Using the Storax/Sedan test estimated coupling efficiency of 2.3\% (see Section \ref{sec:storax}), we would need a total NED yield of $1.24/0.023=54$ Mt for the comet case, and $7.61/0.023=331$ Mt for the asteroid case to completely disassemble and eject the fragments with 5 m/s speed.
\renewcommand{\arraystretch}{2}
\begin{table}[]
    \centering
    \begin{tabular}{|c||c||c|}
         \hline\bf{Threat} & \bf{Comet} & \bf{Asteroid} \\
         \hline\hline Diameter (km) & $10$ & $10$ \\
         \hline Speed (km/s) & $40$ & $20$ \\
         \hline Density (kg/m$^3$) & $650$ & $2600$ \\
         \hline  Mass (kg) & $3.4\times10^{14}$ & $1.36\times10^{15}$ \\
         \hline Gravitational Binding Energy (Mt) & $0.22$ & $3.54$ \\
         \hline Escape Speed (m/s) & $3.0$ & $6.0$ \\
         \hline Initial Fragment Speed (m/s) & $5.8$ & $7.8$ \\ 
         \hline Fragment Kinetic Energy at $\infty$ (Mt) & $1.02$ & $4.07$ \\
         \hline Total Disruption Energy (Mt) & $1.24$ & $7.61$ \\
         \hline NED Mass at 4 Mt/t (ton) & $0.31$ & $1.90$ \\ 
         \hline Delivered Mass at 2.3\% Efficiency (ton) & $13$ & $83$ \\
         \hline 
    \end{tabular}
    \vspace{2.5mm}
    \caption{Comet and asteroid cases with 10 km diameter one month intercept. Required energy and NED mass needed in penetrators to mitigate threat with 5 m/s fragment speed far from bolide. Actual coupled mechanical energy is relatively small, being 1.24 and 7.61 Mt for the comet and asteroid cases respectively. Energy coupling based on estimate from Sedan nuclear test. The NED energy to be delivered to the threat is based on the Sedan test and is 54 and 330 Mt for the comet and asteroid cases respectively with an NED mass of 13 and 83 mt delivered based on a 4 Mt/t NED yield (typical of the best current stockpile weapons). Note that the total mass in a full system would be larger as it must include the penetrator masses as well as guidance and control systems. Required delivered masses are possible with either larger numbers of existing boosters that achieve positive $C_3$ or smaller numbers (including unity) with the next generation of heavy lift positive $C_3$ boosters including SLS and an orbital refueled Starship.}
    \label{tab:comparisontable}
\end{table}

\subsection{Nuclear Penetrators}

While there is no known threat in the foreseeable future from such a large asteroid or comet, it is useful to ponder the question of whether the PI method could mitigate a large existential threat. Due to the $d^5$ dependence of the gravitational binding energy and the $v^2 d^3$ dependence of the disassembly kinetic energy on diameter, as shown, the gravitational binding energy begins to dominate the disassembly energy budget for a $v\sim1$ m/s disruption. For the comet case, with a 5 m/s disruption (needed for fragments to miss the Earth at one month intercept) and a 10 km diameter, density 0.65 g/cm$^3$, the total energy required is roughly 1.2 Mt, while for a 1 km diameter the energy is only about 1.0 kt for a 5 m/s disruption, and only 43 t at 1 m/s disruption. For comparison, the asteroid case with a 10 km diameter, 5 m/s disruption (needed for fragments to miss the Earth at one month intercept), and density 2.6 g/cm$^3$, the total energy required is roughly 7.6 Mt, while for for comparison, for a smaller 1 km diameter asteroid, the energy is only about 4.1 kt for a 5 m/s disruption, and only 200 t at 1 m/s disruption. The mitigation for the larger 10 km diameter cases calls for a nuclear disruption, which is possible, but the details become critical. NED penetrators using modern thermonuclear penetrators, such as the B61-11 or (cancelled) W61 NED, are highly problematic due to the extremely high ``g'' loading upon penetration, while pure fission NED’s may be a better choice, particularly those with a nuclear artillery heritage. However, the problem with fission only NED’s is their low yield per mass compared to thermonuclear NED’s. Another issue is partly political in that NED design is possible, but testing is not due to the Comprehensive Test Ban Treaty (CTBT) that bans all nuclear weapons tests that have nuclear yield. This could always change, but currently this would limit NED options to the existing stockpile or to the design (but not testing) of NED’s that do not require additional testing to ensure their performance. In any realistic scenario of an existential threat, presumably logic would prevail, at least one would hope. If we are confined to the existing stockpile or new untested devices which are ``guaranteed'' to work in full speed penetrators, then pure fission devices are possible. If we use, for example, the (cancelled) W82 pure fission NED (2 kt yield with 43 kg mass) as a conservative example, this gives a yield per mass of 47 t(TNT)/kg. Note that this is far less ($\sim130\times$) than the Taylor limit for thermonuclear devices of 6 kt/kg. Thermonuclear devices are far more mass efficient and would be desired but require a development program to allow for high ``g'' loading penetrators. We discuss this further below.

% \begin{figure}
%     \centering
%     \includegraphics[width=0.5\textwidth]{Figures/Escape Speed vs Diameter- 1 to 15km-linear.png}
%     \caption{Escape speed vs. diameter vs. density.}
%     \label{fig:escapespeed}
% \end{figure}

\subsection{Small Fission-Only NED Penetrators}

At 47 t(TNT)/kg for the W82, we would still have about $250\times$ larger energy than purely kinetic or chemical explosive impactors of the same mass for a 40 km/s closing speed. If we could couple all of the W82 energy into disruption of the 10 km comet case (1.24 Mt minimum mechanical coupled energy needed for 5 m/s disruption including gravitational binding), we would require at least 620 W82 NED’s with a total mass of 27 tons. This mass is compatible with a single SLS, but we are assuming unity coupling efficiency of NED yield into gravitational de-binding and kinetic energy of the fragments, which is unreasonably optimistic as discussed below.

For the asteroid case, we would need at least 7.6 Mt of input energy for our short intercept time (1 month) scenario, which is 6.1 times larger than the energy needed for the comet nucleus case. This would require 165 tons of fission-only interceptor mass.

This pushes us to thermonuclear NED penetrators. If we assume 4 kt/kg (4 Mt/t) for currently stockpiled high efficiency thermonuclear NED’s, as discussed below, we would need 13 tons of delivered NED capacity for the comet case and 83 tons for the asteroid case, both assuming the 2.3\% coupling efficiency discussed below. This mass delivery, while large, is within the realm of feasibility for the next generation of boosters, or a large number of currently existing boosters. In addition, since each NED penetrator is relative low mass, delivery can be via multiple booster. The NED mass estimate does not include the targeting, guidance, and passive/conventional penetrators needed to ``clear the way'' to  allow the NED penetrators to be used. A more detailed analysis of the coupling details is clearly required to better estimate the delivered mass required. 

Note that above about 40km diameter, there is not sufficient explosive energy in the entire world’s current nuclear arsenal to even gravitationally de-bind an asteroid-density bolide. Something to ponder.

\subsection{Thermonuclear NED Penetrators}

The use of well-developed thermonuclear NED’s for ICBM delivery is an option to explore since this reduces the delivered NED mass by a factor of about 100 compared to pure fission NED’s. While currently thermonuclear NED’s are not suitable due to the high $g$-loading upon penetration, a possible path to explore is the use of sequential penetrators to bore a path (``hole drilling'') into the bolide such that subsequent NED’s could enter and not experience the extremely high $g$-loading of the prior penetrators. Detailed simulations and ground testing would need to be done to explore the feasibility of this approach. The use of modest yield NED’s was explored in detail for Operation Plowshare in the US and related programs in the Soviet Union with a number of nuclear tests related to peaceful use capability. Related information from the large number of underground nuclear tests done throughout the 1960’s and 1970’s is also applicable, even though the critical issue of high-$g$ testing is not nearly as extensive, though some data from NED ``bunker busters'' are relevant. As an example, the core thermonuclear physics package in the B61-11 or W61 NED has a yield of about 340 kt, with a mass of about 80 kg ($\sim4$ kt/kg), with a diameter ($\sim30$ cm), small enough to be considered for a penetrator. If such a device could be delivered at low enough $g$ using sequential penetration, then this would be lower mass ($\sim80\times$) than a W82 fission only NED. Note that the numbers for achieved yield-per-mass are slightly less than the Taylor limit (6 kt/kg). A detailed system analysis would have to be done including the possible extra mass of sequential penetrators, targeting electronics, and control thrusters, etc. Verification of robust disruption and fragment tracking would be critical and could be done from both ground and small secondary observation payloads. In such an existential threat scenario, multiple independent interceptors would be necessary as backup, as well as to target large fragments that might still pose a threat. As discussed in our PI paper, residual fragments of order 100 m or less could be dealt with via non-nuclear means with intercept times of less than one day prior to impact if required. A combination of the conventional non-nuclear PI along with the NED PI for existential threats as we have outlined here would make for an extremely robust planetary defense system capable of mitigating virtually any threat. 

\subsection{Ground and Space Penetrator Testing}

Full nuclear testing is not required for NED penetrators if it can be shown that the penetrator $g$-loading is within the limits of the existing NED's. Testing ``instrumented only'' penetrators including sequential penetrator modes can be done at lower speeds on the ground, and full speed ($\sim16$ km/s) testing could be done in space using counter-orbiting LEO-based systems. Non-nuclear penetrator testing using the moon is also an option to be considered. These types of tests need to be done even if NED’s are not used since the penetration depth and energy coupling with passive and chemical explosive penetrators is required. The idea of using sequential penetrators also is a natural area to explore independent of NED’s.

\subsection{Penetrator Methodology -- Onion Peeling}

The same methodology used with the non-nuclear approach as outlined in our PI paper is used here with nuclear penetrators, namely that the penetrators are arranged to impact and detonate such that the outer layers of the target are ejected first, and then successive waves of penetrators work inward to peel away from the outside to the center. We refer to this as ``onion peeling.'' The one primary difference between the nuclear and non-nuclear case is that achieving small fragment size is not critical with the nuclear case as the goal is to spread the fragments out so they miss the Earth completely and the Earth’s atmosphere is not used as a shield as it is in the terminal defense mode of PI, as discussed in our previous paper. 

\subsection{Can't We Just Vaporize the Threat with Enough Nuclear Weapons?}

It is worth pondering this thought. What does it mean to evaporate something? It means to reduce the object to individual free molecular form. When we boil water or evaporate metals we are separating the molecules from each other. How much energy does this take for the types of objects we are considering? Typical binding energy ($\varepsilon$) between molecules is of order $\varepsilon=1$ eV. In one mole, this energy is $E=N_A\varepsilon=9.6\times10^4$ J, or about 100 kJ/mol. For reference, water has a heat of vaporization (liquid to gas) of 41 kJ/mol $\sim2.5$ MJ/kg, while the heat of fusion (solid to liquid) of water is 6 kJ/mol. The heat of fusion is generally about $10\times$ smaller than the heat of vaporization for materials and thus the total energy to go from the solid to the gas phase is dominated by the heat of vaporization. As an example, the total heat required for water from solid to gas is close to 50 kJ/mol (solid to gas) $\sim3$ MJ/kg or about 1/2 of our estimate of 100 kJ/mol. In our papers on directed energy planetary defense, we study the thermophysical properties of a variety of relevant substances for asteroid and comet nuclei. Typical heat of relevant materials in going from solid to gas phase is 1--10 MJ/kg, similar to that of water. We can now estimate the total energy required to vaporize a relevant target. With our case of 10 km diameter and 2.6g/cm$^3$, we have a mass ($m=\rho\pi d^3/6$) of about $m=1.4\times10^{15}$ kg. If we assume the lower heat of 1 MJ/kg to go from solid to gas, we get a total energy to vaporize our target of $1.4\times10^{21}$ J.  With 1 Mt $\sim4.2\times10^{15}$ J, this yields $3.2\times10^5$ Mt, or 320 Gt, which is about $50\times$ larger than the entire worlds arsenal of nuclear weapons. So, definitively NO, we cannot vaporize our target with nuclear weapons. Looking at Figure \ref{fig:disassemblyenergy}, we see this is a reasonable estimate as the speed of molecular ejection for vaporization of common materials is roughly 1 km/s and thus the energy required is completely dominated by the kinetic energy of the material and is of order 200 Gt for 1 km/s ($100\times$ 100 m/s) energy disruption. Note that the disruption energy is independent of the fragment size whether a molecule or a city in size. Note that even if we could vaporize the target and it did not have enough time for the vapor cloud to spread out, such as in a terminal defense mode, and we allowed the vaporized cloud to hit the Earth’s atmosphere, the resulting damage from absorbing the entire bolide energy in our atmosphere would cause significant heating with possible catastrophic results. See the PI paper for a more detailed discussion of this.

\begin{figure}
    \centering
    \includegraphics[width=0.6\textwidth]{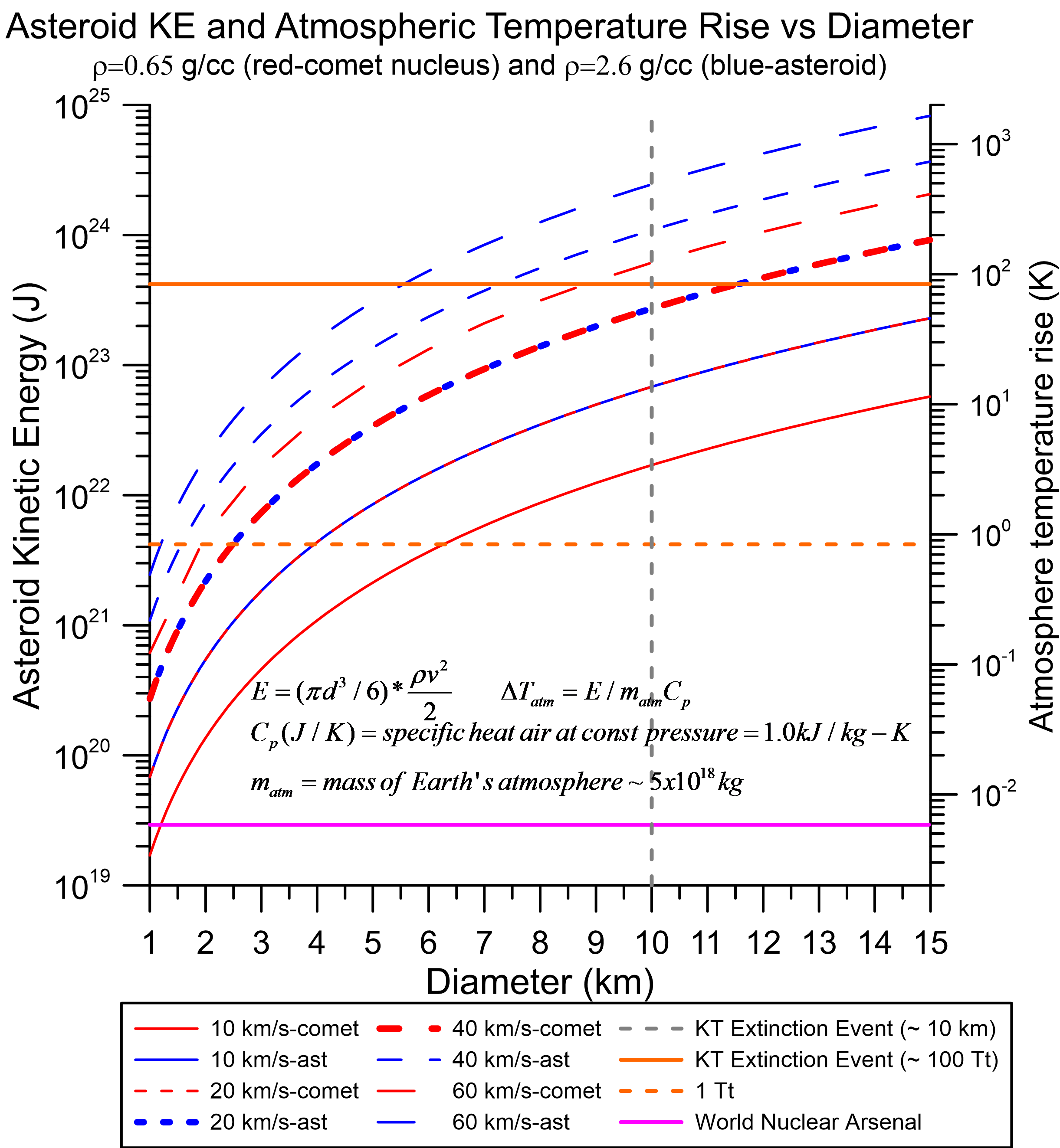}
    \caption{Average atmospheric temperature rise assuming homogeneous distribution of the total kinetic energy of the bolide fragments. The scale size of the fragments is largely irrelevant here though diameters above about 80 m will penetrate to the ground.}
    \label{fig:temprise}
\end{figure}

\subsection{Can't We Just Fragment the Threat Into Small Fragments and Accept the Hit, Like in PI?}

Unfortunately, while the atmosphere works extremely well as a shield for fragmented bolides less than about 1 km in diameter, it does not have large enough heat capacity to accept the total amount of energy from a large bolide of the 10 km class this paper is focusing on. Unlike the case we made in the previous PI papers that focused on non-existential threats (less than a few km diameter), the situation is quite different for large existential threats. It becomes a question of atmospheric heating. The total amount of kinetic energy in a 10 km diameter bolide is too much for our atmosphere to absorb without catastrophic results no matter how small we fragment the parent bolide \textit{if} we accept the hit from all the fragments. As is seen in Figure \ref{fig:temprise}, the atmospheric temperature rise assuming uniform distribution of the bolide fragments kinetic energy (best case scenario) is catastrophic for large bolides like our 10km threat. Note that the average temperature \textit{rise} for a 10 km diameter, 40 km/s comet with density 0.65 g/cm$^3$ is roughly 55 C, with local heating much higher. The average temperature \textit{rise} for a 10 km diameter, 20 km/s asteroid with density 2.6 g/cm$^3$ is the same as the comet as the kinetic energy is the same. This would destroy much of life on the surface of the Earth, though underground and underwater life could survive, as it did in the KT event that killed the dinosaurs. A possible mode of survival for humanity in such a case of terminal defense fragmentation and acceptance of the hit, like in our previous paper, is for humanity to go underground during the event (assuming fragmentation), but the resulting damage to the Earth’s surface ecosystem would be truly catastrophic.

It is worth noting that the total kinetic energy of all the molecules in the atmosphere is comparable to the kinetic energy of our 10 km threat, whether comet or asteroid. This is a sobering statement. A larger example at 15 km diameter and 60 km/s speed would leave parts of the atmosphere significantly ionized. In a ``real event'' of fragmentation to the scales discussed in our previous PI paper ($\sim10$ m diameter fragments) the local heating of the air near the fragments would cause local ionization. The atmosphere in any real scenario would be locally ``super-heated'' and there would be a very inhomogeneous temperature distribution which would cause significant opacity in the ionized air regions as well as the fragments that would subsequently radiate their energy and cause massive surface fires even without a ground hit of any fragment. In addition, the radiated heat from the bolide fragments would also cause massive fires through thermal radiation. This is discussed in detail in our previous PI paper. 

\begin{figure}
    \centering
    \includegraphics[width=0.6\textwidth]{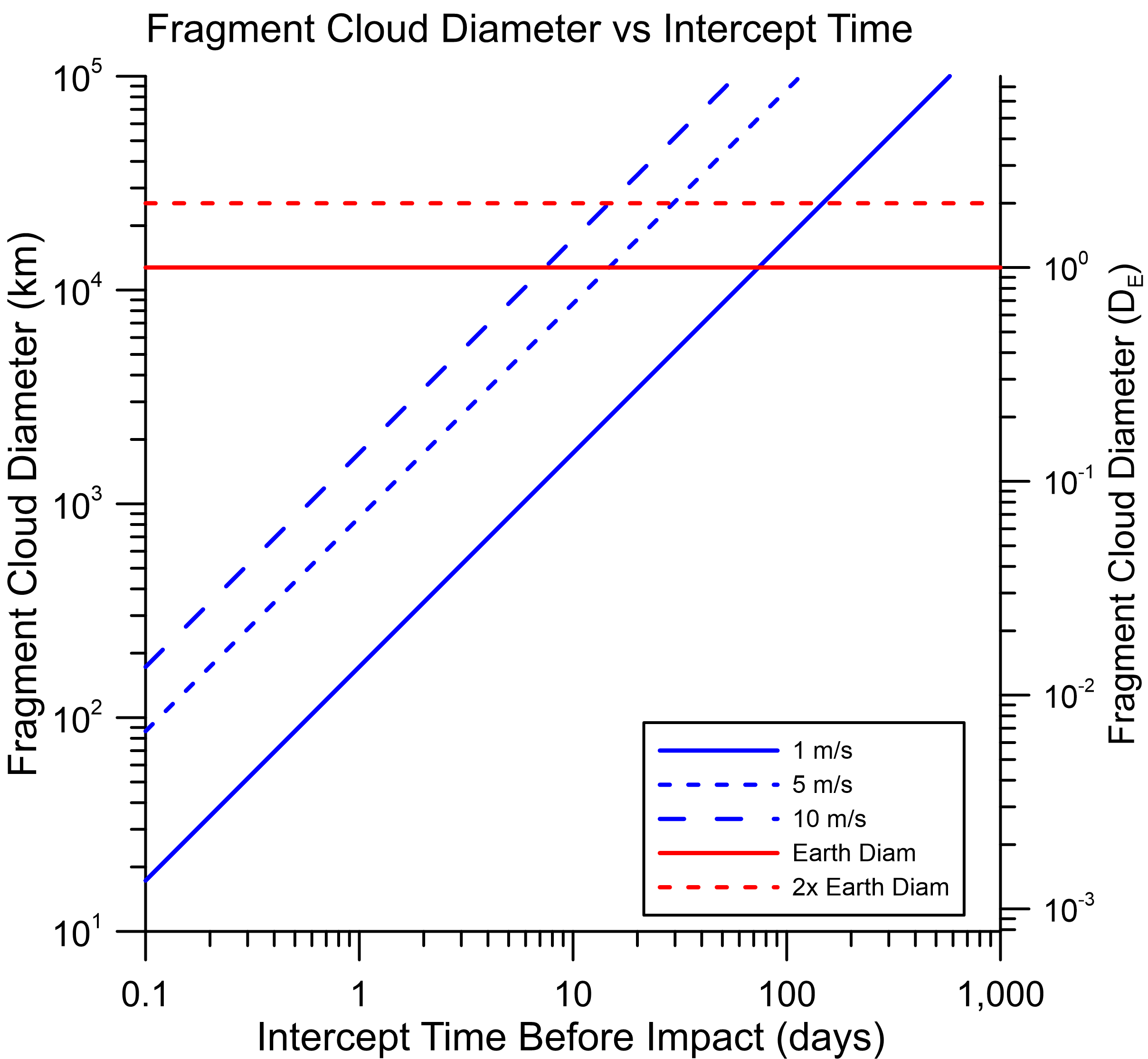}
    \caption{Fragment cloud diameter vs. time before impact assuming the fragments have a given average speed relative to the parent asteroid center of mass. After 75 days at 1m/s fragment speed, the cloud diameter is larger than the Earth and thus the fragments largely miss the Earth depending on the initial angle of attack of the parent bolide. For the worst case of the angle of attack being 90 degrees (edge of Earth), then the cloud diameter needs to be $2\times$ Earth diameter, giving 150 days at 1 m/s, or 30 days at 5 m/s.}
    \label{fig:clouddiameter}
\end{figure}

\subsection{Intercept Time Prior to Impact}

With a goal of intercepting, fragmenting, and dispersing the fragment cloud with sufficient dispersal speed to have the fragments miss the Earth, we need the product of the intercept time prior to impact times the fragment speed to exceed the diameter of the Earth in the worst case of an unmitigated threat at the ``edge'' of the Earth (bolide velocity vector normal to local impact normal vector). The product of fragment speed and time to impact after intercept is the critical metric. For a 1 m/s transverse fragment speed and 75 days' time, the fragment has moved to equal the Earth’s radius, while at 2 m/s and 75 days, the fragment cloud radius has spread a distance of the Earth’s diameter. This is summarized in Figure \ref{fig:clouddiameter} for 1, 5, and 10 m/s fragment speeds. As described earlier, for a 10 km bolide the gravitational binding energy dominates for fragment speeds less than about 1 m/s, but for the desired 5 m/s the fragment kinetic energy dominates the energy budget. At 5 m/s (at infinity), fragment speed and one-month intercept prior to impact the fragments will miss the Earth. Shorter intercept times or larger safety margin are viable for faster fragment speeds, which require greater energy input. This is a trade space for any threat. 

For reference, at 40 km/s an intercept at one month prior to impact of the comet would be at a distance of about 0.69 AU, or roughly $273\times$ the distance to the moon, while for the 20 km/s asteroid, the distance would be half this (where 1 lunar distance (LD) $\sim380,000$ km).

\subsection{Interceptor Speed and Required Launch Time Prior to Impact}

To intercept the target in sufficient time prior to impact, the interceptor must be launched early enough so the interceptor time to target is appropriate. The ratio of the required intercept time prior to impact divided by the time impact at the moment of launch is shown. We assume the interceptor reaches full speed in a short time compared to the intercept time to impact. As shown, the interceptor (spacecraft) speed is extremely important, especially for fast moving bolides. As shown in the next section, this places a severe requirement on launch vehicle capability for the putative threat case we have posited of 10 km diameter, 0.65 g/cm$^3$, 40 km/s speed for the comet and 10 km, 2.6 g/cm$^3$, 20 km/s speed for the asteroid, both with 6 months warning.  As an example, at 10 km/s interceptor speed for the comet threat (40 km/s), the time from launch to impact is $5\times$ longer than the intercept time to impact. For the asteroid threat (20 km/s) with a 5 km/s interceptor, the time from launch to impact is also $5\times$ longer than the intercept time to impact. Thus, if we had 6 months warning and launched at 5 months prior to impact, the intercept would occur one month prior to impact. For the fragments to miss the Earth with one month prior to impact interception, we would need about 5 m/s fragment speed at infinity. For our hypothetical case, the minimum injected energy would be about 1.24 Mt for the comet case and 7.61 Mt for the asteroid case. 

As we will see in the next section on current launcher capability, 13 mt of mass delivered (comet case) at 10 km/s is possible with a single SLS Block 2B. For the asteroid case with 5 km/s interceptor, the required mass delivery of 83 mt would require multiple SLS Block 2B launches. A single SpaceX Starship refueled in orbit may be another viable option. It is interesting to note that current human capability is ``right on the edge'' of being viable to take on the extreme threat scenario we have outlined. Note that we have assumed such an extreme case as an example to show that even with such a short notice, existential threat, we could potentially defend ourselves. This, in itself, is a remarkable statement. 

% \begin{figure}
%     \centering
%     \includegraphics[width=0.5\textwidth]{Figures/Time after intercept vs Spacecraft speed-b.png}
%     \caption{Ratio of $\tau_\textrm{int}/\tau_0$ vs. spacecraft speed and asteroid speed where $\tau_\textrm{int}$ is the time to hit the Earth after intercept and $\tau_0$ is the time to hit the Earth after launch. Note 10 km/s $\sim0.17$ AU/month.}
%     \label{fig:timetoimpact}
% \end{figure}

\begin{figure}
    \centering
    \includegraphics[width=0.5\textwidth]{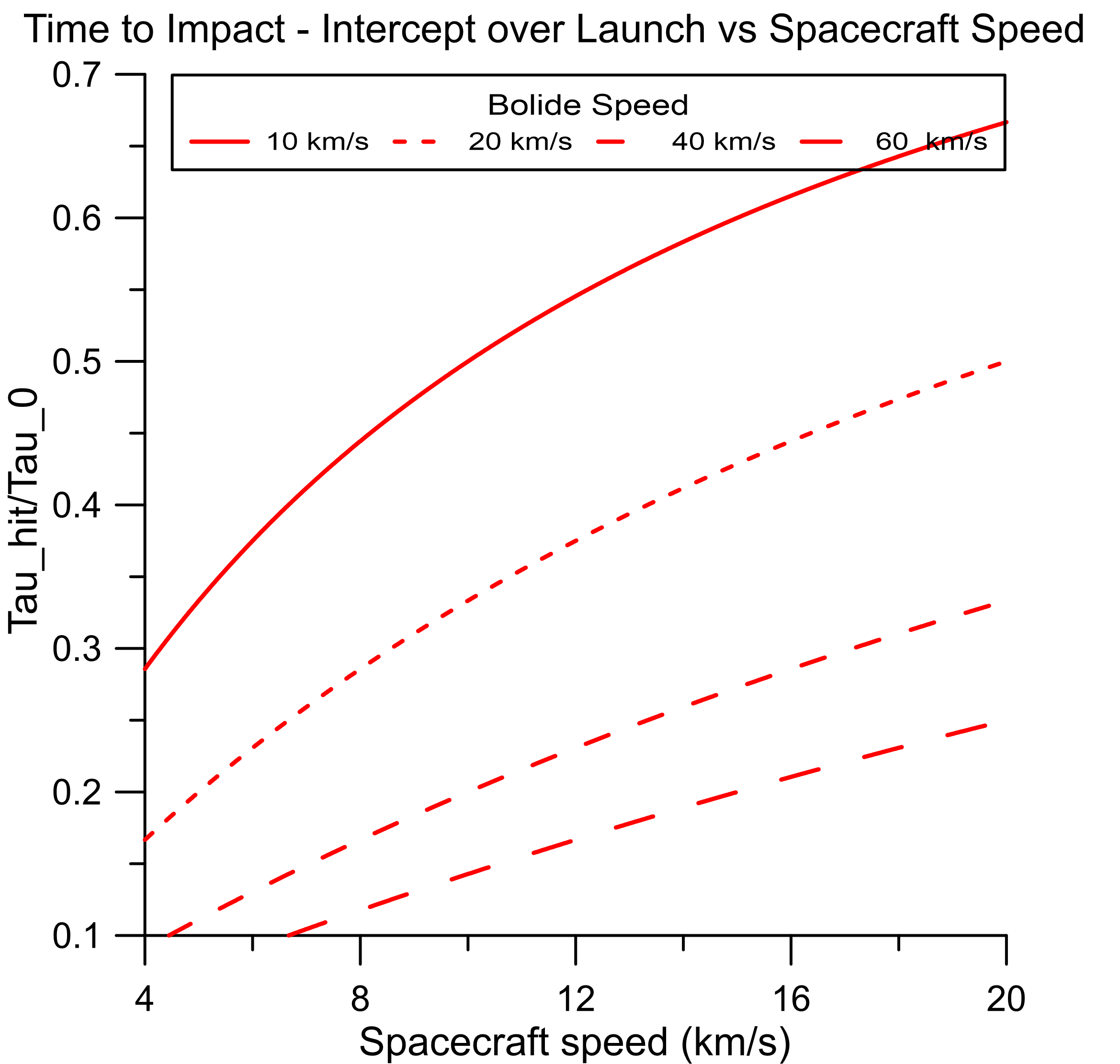}
    \includegraphics[width=0.7\textwidth]{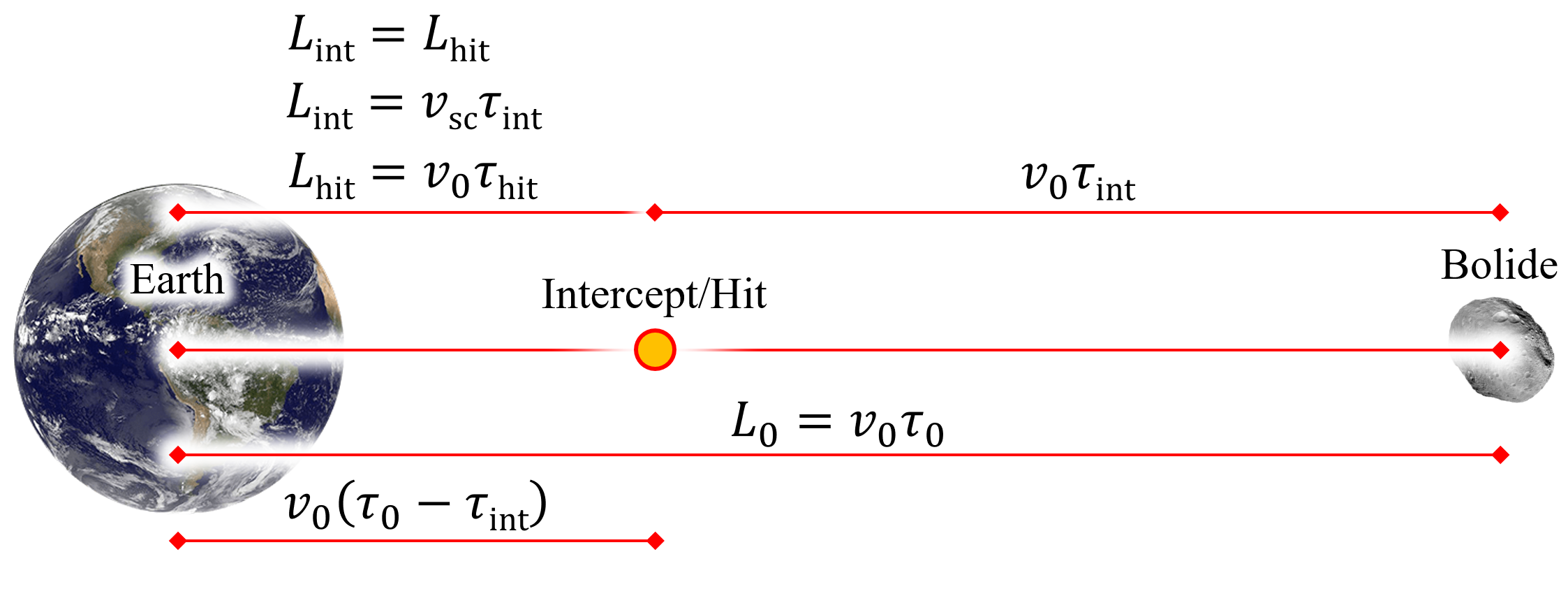}
    \caption{(Top) Ratio of $\tau_\textrm{int}/\tau_0$ vs. spacecraft speed and asteroid speed where $\tau_\textrm{int}$ is the time to hit the Earth after intercept and $\tau_0$ is the time to hit the Earth after launch. Note 10 km/s $\sim0.17$ AU/month. (Bottom) Relationship between intercept, spacecraft speed, and initial bolide position. For simplicity, this ignores the complexity of the actual orbital dynamics which are case specific.}
    \label{fig:intercept}
\end{figure}

As we will see in the next section on current launcher capability, 25 mt delivered at 10 km/s is not currently quite viable with any currently proposed single SLS variant (Earth-launched), though two SLS block 2B could conceivably work. A single ``refueled-in-orbit'' SpaceX Starship may be another viable option. It is interesting to note that current human capability is right on the edge of being viable to take on the extreme threat scenario we have outlined. Note that we have assumed such an extreme case as an example to show that even with short notice existential threats we are close to defending ourselves. This in itself is a remarkable statement. 

\begin{align}
    \begin{split}
        v_0 &= \textrm{target speed relative to Earth,} \\
        v_\textrm{sc} &= \textrm{spacecraft speed relative to Earth,} \\
        L_\textrm{int} &= L_\textrm{hit} = \textrm{distance to target at intercept/hit,} \\
        L_0 &= \textrm{distance to target at launch (1 AU $\sim1.5\times10^{11}$ m $\sim395$ LD),} \\
        \tau_\textrm{int} &= \textrm{time for spacecraft to intercept starting from $t=0$ at launch,} \\
        \tau_0 &= \textrm{time for target to collide with Earth starting from $t=0$ at launch,} \\
        \tau_\textrm{hit} &= \textrm{time for target to collide with Earth from $t=\tau_\textrm{int}$.}
    \end{split}
\end{align}

\begin{equation}
    \tau_0=\frac{L_0}{v_0}, \hspace{3mm} \tau_\textrm{int}=\frac{L_\textrm{int}}{v_\textrm{sc}}, \hspace{3mm} \tau_\textrm{hit}=\frac{L_\textrm{hit}}{v_0}=\frac{L_\textrm{int}}{v_0}.
\end{equation}
\begin{equation}
    L_\textrm{int}=v_\textrm{sc}\mathop{\tau_\textrm{int}}=L_0-v_0\mathop{\tau_\textrm{int}}=v_0\left(\tau_0-\tau_\textrm{int}\right).
\end{equation}
\begin{equation}
    \tau\textrm{int}=\frac{L_0}{v_0+v_\textrm{sc}}=\frac{\tau_0\mathop{v_0}}{v_0+v_\textrm{sc}}, \hspace{3mm} \tau_\textrm{hit}=\frac{L_\textrm{hit}}{v_0}=\tau_0-\tau_\textrm{int}.
\end{equation}
\begin{equation}
    \frac{\tau_\textrm{hit}}{\tau_0}=\frac{v_\textrm{sc}}{v_0+v_\textrm{sc}}, \hspace{3mm} \frac{\tau_\textrm{int}}{\tau_0}=\frac{v_0}{v_0+v_\textrm{sc}}, \hspace{3mm} \tau_\textrm{hit}+\tau_\textrm{int}=\tau_0, \hspace{3mm} L_\textrm{hit}=v_0\mathop{\tau_\textrm{hit}}=v_\textrm{sc}\mathop{\tau_\textrm{int}}=L_\textrm{int}.
\end{equation}
\begin{equation}
    \mathop{\textrm{Ex:}} v_0=\mathop{40}\textrm{km/s}, \hspace{1mm} v_\textrm{sc}=\mathop{10}\textrm{km/s} \rightarrow \frac{\tau_\textrm{hit}}{\tau_0}=\mathop{0.2}, \hspace{1mm}\textrm{if} \hspace{1mm} \tau_0=\mathop{5}\textrm{months}, \hspace{1mm} \tau_\textrm{int}=\mathop{4}\textrm{months}, \hspace{1mm} L_\textrm{int}\sim\mathop{0.69}\textrm{AU}\sim \mathop{273}\textrm{LD.}
\end{equation}
\begin{equation}
    \mathop{\textrm{Ex:}} v_0=\mathop{20}\textrm{km/s}, \hspace{1mm} v_\textrm{sc}=\mathop{5}\textrm{km/s} \rightarrow \frac{\tau_\textrm{hit}}{\tau_0}=\mathop{0.2}, \hspace{1mm}\textrm{if} \hspace{1mm} \tau_0=\mathop{5}\textrm{months}, \hspace{1mm} \tau_\textrm{int}=\mathop{4}\textrm{months}, \hspace{1mm} L_\textrm{int}\sim\mathop{0.35}\textrm{AU}\sim \mathop{136}\textrm{LD.}
\end{equation}

\subsection{Trade and Optimization Space}

Note that there is a tight coupling between warning time, target speed, and NED energy requirements, and thus delivered mass and the launcher requirements. In our extreme threat case, we have deliberately chosen a very difficult threat to mitigate. Our threat at 40 km/s speed is in a reasonable range of comet speeds and 20 km/s is also a reasonable speed for an asteroid. In either case (asteroid or comet), this threat would be a ``planet killer'' if not mitigated. 

\subsection{Energy Injection Requirement Scaling with Warning Time}

For our case of a 10 km diameter comet, the gravitational binding energy alone is about 0.22 Mt of energy, and though longer warning time decreases the required fragment speed, the fragment speed is inversely proportional to the warning time, and thus the required fragment kinetic energy decreases inversely proportional to the square of the warning time, but the gravitational binding energy does not change with warning time. Thus, the minimum energy we must inject is the gravitational binding energy for infinite warning time. We still need to consider the other issues we have mentioned, such as energy coupling efficiency, penetrator masses, guidance and control, safety factor, etc. Thus, the difference between one month intercept and infinite intercept time is only about a factor of $\sim6\times$ in energy delivery required for the comet case. For the asteroid case, the gravitational binding energy is 3.54 Mt, while the 5 m/s fragment kinetic energy is 4.07 Mt, and thus the difference between one month intercept and infinite intercept time is only about a factor of $\sim2\times$ in energy delivery required for the asteroid case.

\subsection{The Problem is Not the Energy Required, but the Coupling Efficiency and Spatial Distribution}

For our case of a 10 km diameter comet with 6-month warning and an intercept one month prior to impact, the gravitational binding energy is 0.22 Mt of energy and an additional 1.02 Mt to impart the required 5 m/s average fragment speed gives the desired total of about 1.24 Mt fully coupled energy injection. It is feasible to deliver 1.24 Mt of energy in NED’s, but the critical issue is the efficient coupling of NED energy released and the disruption energy required (gravitational binding energy + fragment kinetic energy). While an airburst NED couples $\sim50$\% of its energy into the acoustical blast wave and a well-tamped chemical explosion couples about 50\% of its chemical energy into mechanical motion, it will be a challenge to efficiently couple the NED energy into the bolide and spatially distribute it to meet our needs. This is an area that requires extensive simulation and comparison to the previous work in Project Plowshare and the many underground tests, as well as in detailed studies and some tests of NED Earth penetrators for ``deep bunker busting.'' The NAS 2005 study is particularly relevant \cite{national_research_council_effects_2005}

\subsection{Storax Sedan Project Plowshare NED Test}
\label{sec:storax}

\begin{figure}
\centering
         %\centering
         \includegraphics[width=0.4\textwidth]{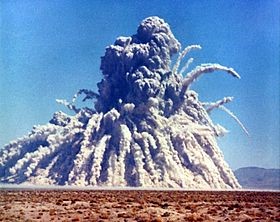}
         %\caption{Caption}
         %\label{fig:my_label}
         %\centering
         \includegraphics[width=0.4\textwidth]{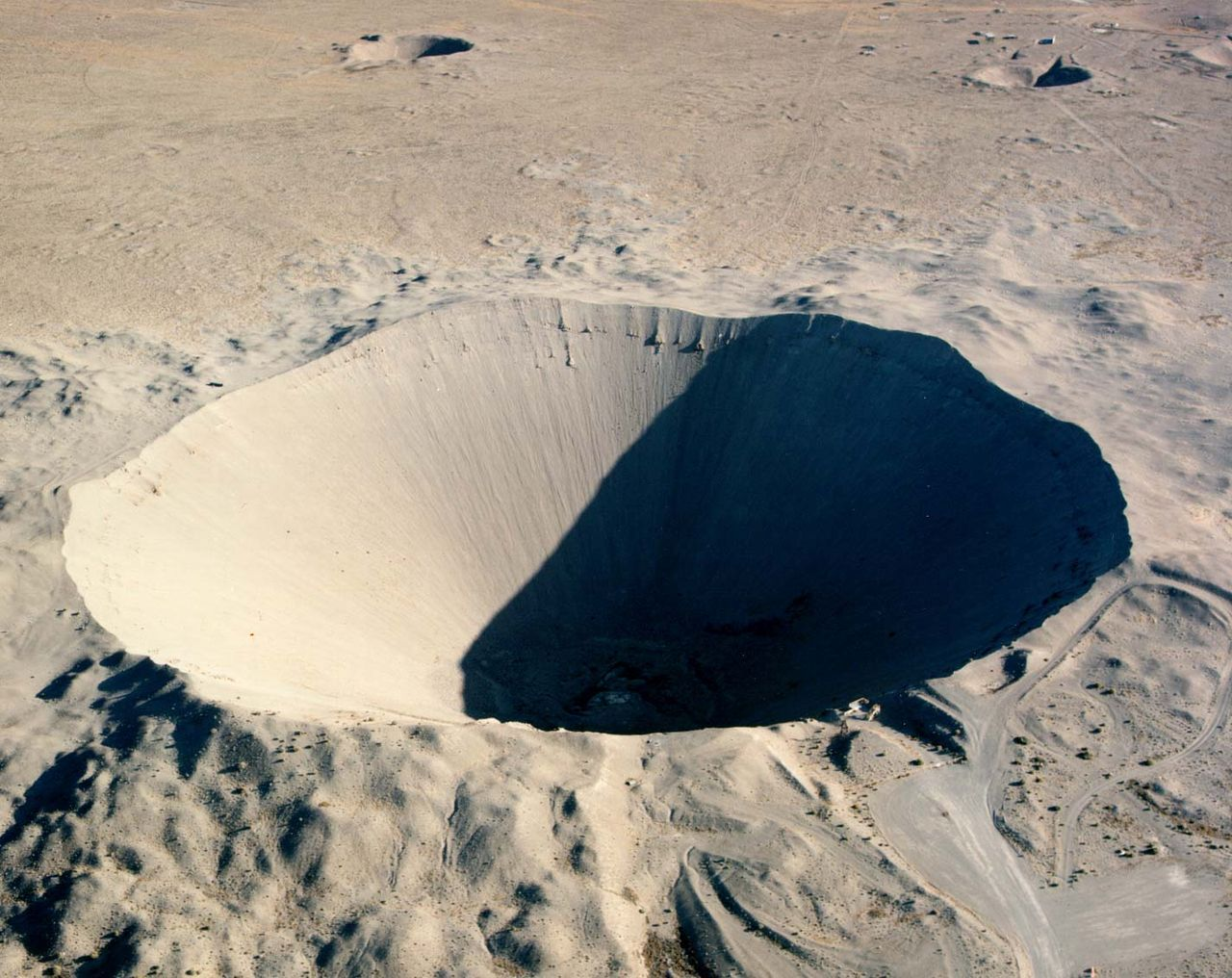}
         %\caption{Caption}
         %\label{fig:my_label}
     \caption{July 6, 1962 underground Storax Sedan (Project Plowshare) test (104 kt). Left:  debris cloud. Horizontal image scale is $\sim500$ m. Right: 390 m diameter $\times$ 100 m deep crater formed.}
     \label{fig:plowshare}
\end{figure}

One relevant example of energy coupling efficiency is the July 6, 1962 Storax Sedan test of a 104 kt (435 TJ) yield NED (30\% fission/70\% fusion), which is pictured in Figure \ref{fig:plowshare}. This earlier generation NED had a mass of about 212 kg with a yield per mass of about 0.5 kt/kg, about a factor of 6--10$\times$ lower than the highest yield per mass devices currently possible. A 194 m deep shaft was drilled into the Nevada area 10 Yucca Flat dessert alluvium soil (silt, sand, clay, gravel mixture). The blast displaced around $1.12\times10^7$ mt ($1.12\times10^{10}$ kg) of soil and created a crater about 390 m in diameter and 100 m deep. It lifted the Earth dome about 90 m high before venting. There was a horizontal flow of debris about 8 km in diameter similar to a pyroclastic base surge of a volcanic eruption. The resulting crater is referred to as the Sedan Crater and is the largest human made crater in the United States.  Note that at the depth of 194 m, the blast was not fully contained, but is still a good reference point for this paper. If we assume an average soil density of 2.6 g/cm$^3$, which is typical of alluvium, and that all of the dome was lifted 90 m high and ignore the additional energy release from venting and the 8 km diameter horizontal ``surge flow,'' as well as the resulting magnitude 4.75 earthquake, we get a total mechanical energy of the dome of 9.9 TJ. This gives a (dome only) coupling efficiency of $9.9/435=2.3$\%. If we add in the seismic energy of the magnitude $m=4.75$ earthquake, we get an additional energy of $\mathop{E}\textrm{(J)}=10^{1.5m+4.8}\sim0.84$ TJ, which is a relatively small additional energy correction. It is not clear how much additional energy was in the 8 km diameter horizontal surge flow. We consider the 2.3\% coupling coefficient to be very pessimistic and a lower limit.

This is a highly simplified analysis and likely an under-estimated of the energy of this complex NED-soil interaction, but gives us some sense of the coupling. Note that we are also ignoring the fact that the weapon was detonated at 194 m beneath the surface. Assuming an alluvium density of 2.6 g/cm$^3$, this translates into a ``over-pressure'' due to the Earth’s gravity of $P=\rho gh=4.9$ Mpa $\sim49$ bar. This over-pressure would be negligible for a 10 km bolide for penetration and detonation near the outer layer of the bolide, and hence we can expect a significantly higher coupling efficiency for the bolide than for the Earth. Note that the ejection speed to reach 90 m height is $v=\sqrt{2gh}\sim42$ m/s, or much larger than the nominal 7 m/s needed to reach 5 m/s at infinite distance from our 10 km bolide.  Likely a more deeply ``buried'' penetrator would be able to blow off a much greater amount of material from our target if optimized. Much more work needs to be done in large bolide fragmentation with NED’s to arrive at a better estimate of the coupling coefficient.

Nonetheless, if we use 2.3\% as an estimate of the minimum coupling, then we would conclude that for the comet case the required 1.24 Mt of mechanical energy needed would require a total NED energy release of about 54 Mt. If we assume a yield per mass of 4 mt/t, then this would require a delivery of $\sim13$ mt (83 mt for the asteroid case) of bare NED penetrators. While this is not beyond our ability to conceivably deliver with multiple SLS or Starship’s, it is not a comfortable place to start from. In addition, we still must face the complex issue of delivering the NED’s spatially so that they can couple in a manner to fragment the entire bulk of the bolide as well as the critical issue of the survival of the NED’s within the penetrator during the penetration phase.

\subsection{NED Penetrator Array Size and Multiple Penetrator Waves}

If we assume that the results of the Storax Sedan test are relevant to the dimension of the fragmented and ejected material we need, then we would conclude that a $d=0.5$ km spacing between 100 kt NED penetrators is needed. Again, this is using the existing Storax Sedan test which is a non-optimized case for our ultimate purposes. One delivery scenario is to use successive waves of penetrators working from the outside inward. For an assumed spherical bolide geometry with volume being the relevant metric, then we would conclude as a ROM that we would need about $N=\left(\pi/6\right)\left(10/d\right)^3\sim4000$ of the 100 kt yield NED penetrators for a total of 400 Mt, $8\times$ the total yield (comet -- 54 Mt) or about the total yield (asteroid -- 330 Mt) that we computed in the previous section. This is a highly pessimistic case, but serves as a worst case scenario.

Note that the total number and hence total mass of the NED’s is extremely sensitive to the spacing of the detonations $d$. The total NED mass is roughly consistent with the 110 mt of bare NED’s calculated in the previous section. There is an economy of scale with NED’s where low yield NED’s have lower yield per unit mass. If we take the B61 physics package as an example, this NED has a yield of approximately 340 kt for a rough yield per mass of 4 kt/kg ($\sim85$ kg bare NED).  There is nothing ``magical'' about the Storax Sedan test using a 104 kt NED. It is simply an existing data point. As mentioned, the Storax test is likely a lower limit to the coupling efficiency and that the numbers calculated here are overly pessimistic. Note that 100 kt is probably too small to be optimized for minimum mass. There are also \textit{not} 4,000 of this class NED’s in the US stockpile, though the total yield of 400 Mt for this pessimistic case is well within the US stockpile total yield. Optimization of any real NED based existential threat mitigation would require a careful analysis of existing NED’s, their possible reconfiguration, and the penetrator designs that would house them. This would be a significant task and not consistent with a short term notice and would clearly need to be done long before such an existential threat was detected.

Vastly more analysis and simulations need to be done to develop an optimized penetrator/NED array topology and delivery scenario. The numbers above are a rough (pessimistic) order of magnitude.

\subsection{Residual Radiation Concerns}

In any nuclear mitigation, there would likely be public concern about creating radioactive ``asteroids/comet fragments.'' This is generally not a concern since, by the nature of the system, the fragments miss the Earth. In addition, any residual fragments could be intercepted and fragmented, in which case the remaining fragments are dispersed and burst in the Earth's atmosphere. Since the time from intercept to any residual fragments hitting the Earth’s atmosphere is of order 1 month in our scenario, any highly radioactive, short-lived isotopes would largely have decayed over the intervening month of transit and any residual radiation would be at a low level. Comparing to the total amount of atmospheric nuclear testing done on Earth, the total radiation of the residual fragments is extremely small. An an example, in the Storax Sedan NED test, the crater was found to have low enough levels of radiation that after 7 months, the bottom of the crater could be walked upon with no protective clothing. This should alleviate such radiation concerns for any fragments that may revisit the Earth in the future. Of course, we would prefer to avoid fragments from the initial hit altogether. As a relevant example, after one month the radiation level in the Sedan crater was already down to 500 mR/hr and dropping rapidly.

\begin{figure}[h]
    \centering
    \includegraphics[width=0.6\textwidth]{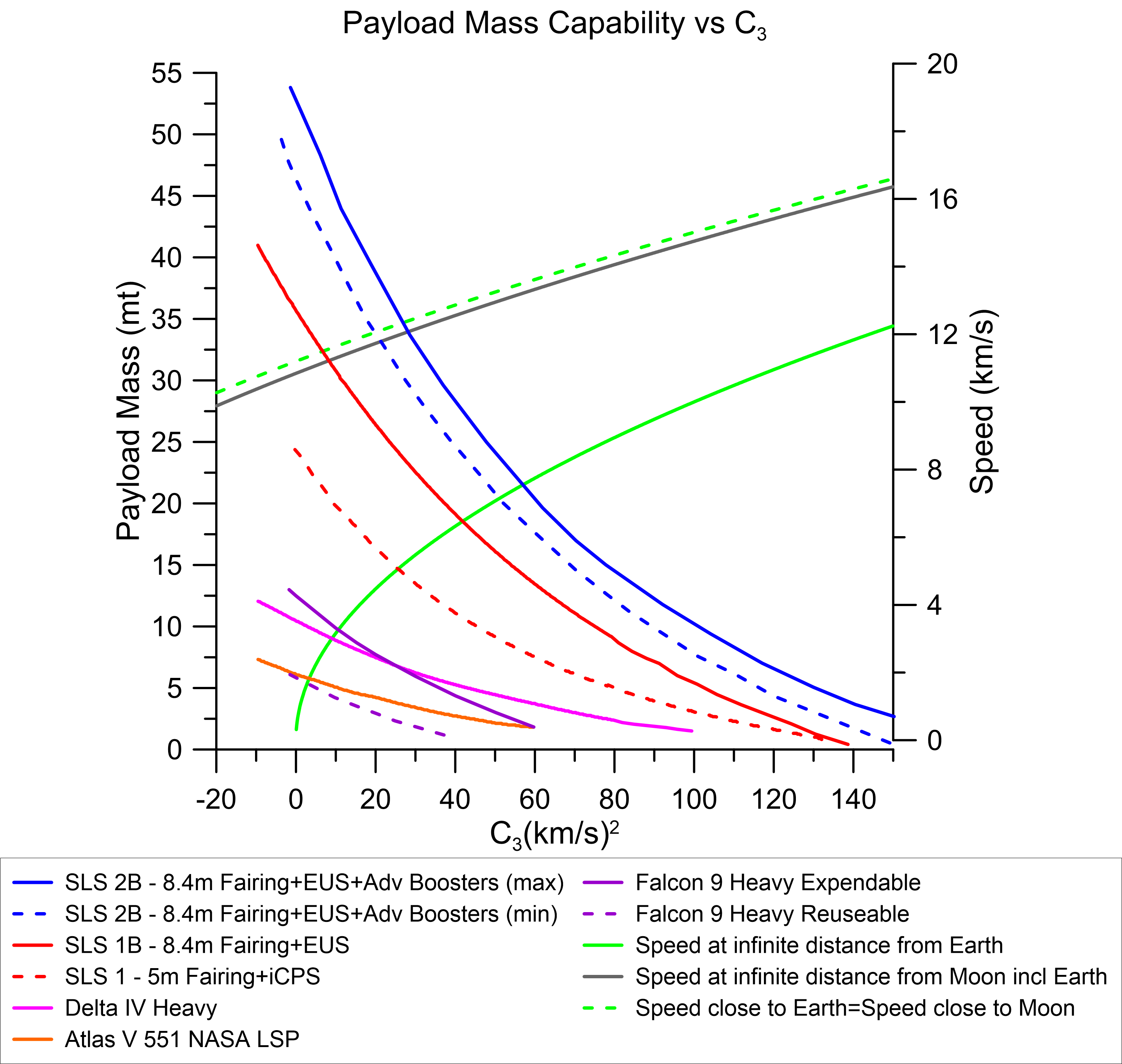}
    \caption{Payload mass and speed for both Earth and lunar launcher vs $C_3$ if the same launcher were used in both an Earth and lunar launch scenario. Note that the payload speed for the same launcher from a lunar launch capability is vastly larger. While it does not make sense to have the same launcher on the Moon for the launchers shown, it does show that lower performance launchers, in particular the use of solid fuel launchers, would become viable on the Moon while they are not viable on the Earth.}
    \label{fig:C3}
\end{figure}

\subsection{Quadratic Fragment Speed Dispersion Issue}

One concern to be studied in detailed simulations is in the fragmentation speed distribution of the resulting fragment dispersal. Since the dispersal energy required (not the gravitational binding energy) is quadratic in the speed of the fragments, it will be important to minimize the mass of fragments that are dispersed at significantly higher speeds than are needed, as the higher speeds take more energy. Small amounts of mass in high speed fragments is not much of a concern, but larger fractions would be. Although higher speeds in the fragments is good in the sense that these fragments ``clear the Earth,'' they also take more energy than is necessary. This will push us to be conservative in the amount of energy input we require and thus additional energy input will be wise.

\subsection{Extremely High Speed Penetrator Interception -- Guidance and Targeting}

We have little experience in locking onto and hitting targets at closing speeds of 10’s of km/s. While we have been relatively successful in developing rapid reaction ICSB interceptors, the speeds, while high, are not in the range of 20--40 km/s-class interception needed for the threats discussed here. This requires a development program which can be tested on both counter-rotating LEO frames (16 km/s closing speed), as well as on ``targets of opportunity'' that arise in asteroids that approach Earth. Any true planetary defense program, even against relatively small targets, would require an extensive development and testing phase to validate its readiness. Focusing on small targets of opportunity or synthetic targets using inert or conventional explosive-filled penetrators would be a part of any such program.

\subsection{Interceptor Capability}

We show some of the various launcher options for the interceptor. For reference, we show two possible launch scenarios using the same launcher. One is on the Earth and the other is on the Moon. With the much lower escape speed from the lunar surface vs the Earth’s surface, there is a significantly higher speed reached far from Earth. The typical way that launchers are characterized for deep space missions which are relevant for our needs is to quote the characteristic energy, known as $C_3$, as a function of the payload capability. The speed obtained far from the Earth or the Moon as appropriate for the launch site is also shown. The problem with lunar launches is that we have no such current capability. In the longer term, there is a significant advantage to orbital-based (LEO or GEO) or lunar-based launchers compared to Earth-based launchers for high mass delivery applications. We show in our ``PI paper'' that lunar/orbital based solid rocket boosters are a viable option if such capability were desired. Clearly, political consideration is important in any such basing option. The SpaceX Starship option is not shown due to the lack of published data. A LEO refueled Starship would be a possible candidate for a high $C_3$, high mass delivery system with possible delivered masses of 100 metric ton (mt). Such a refuel-in-orbit option may avoid some of the political and technical issues associated with LEO/GEO/lunar basing. Depending on the speed of our hypothetical 10 km diameter threat, the use of a single SLS 1B or 2B could be sufficient for the comet threat, while two SLS block 2B or a single SpaceX Starship could suffice for the asteroid threat. 

Note that the required energy, and thus delivered mass, changes the speed for a given launcher (see Figure \ref{fig:C3} $C_3$ vs. mass delivered). Using more interceptors, each of which has lower delivered mass, means they are faster and hence we intercept earlier, which means we need lower fragment speed, which requires less NED mass. This is a part of the trade space that would need to be done for any given threat.

\subsection{Comparison to Comet NEOWISE}

Comet NEOWISE C2020/F3 is a long period comet that was discovered by IR telescope WISE on March 27, 2020 at 2 AU from the Sun, and is a good example of a significant threat that was discovered with relatively short notice prior to perihelion on July 3, 2020 passing about 0.295 AU from the Sun and closest approach to Earth on July 23 at a distance of 0.36 AU ($140\times$ lunar distance). Note that there was only 4 months from discovery to closest Earth approach. With a diameter of approximately 5 km, had it hit the Earth it would have caused catastrophic damage with an impact energy of 10--40 Tt depending on speed. We assume it has a nucleus density of 0.65 g/cm$^3$, as we have assumed for our hypothetical 10 km comet threat. As shown in Figure \ref{fig:disassemblyenergy}, the energy required to gravitationally de-bind and disrupt comet NEOWISE at 5 m/s at infinity would only have been about 0.13 Mt rather than the 1.24 Mt needed for our hypothetical 10 km diameter threat. This $10\times$ reduction would make our mitigation much easier, however the even shorter warning time of four  months for NEOWISE would mean that if comet NEOWISE were on a collision course with the Earth, and if we were to launch three months prior to impact, and if it were also travelling at 40 km/s average speed, and if we were to use similar spacecraft capability to mitigate with 10 km/s speed, then we would have intercepted at $0.2\times3=0.6$ month prior to impact. To have the fragments clear the Earth. we would need the fragment speed at intercept to be larger by $1/0.6\sim1.7\times$, or a fragment speed of about 8.5 m/s instead of 5 m/s we used for our 10 km threat for which we have a 6-month warning and intercepted one month prior to impact. This would push our minimum energy required to about 0.37 Mt to disassemble C2020/F3 and spread the fragments to miss the Earth, much less than our 1.24 Mt minimum needed for our 10 km diameter threat. Note that the gravitational binding energy of comet NEOWISE is only about 6.9 kt and hence it could be taken apart ``quite easily'' compared to our 10 km diameter threats.

NEOWISE is a good example of real short notice (sungrazer comet) large potential threat that did not endanger the Earth, but this emphasizes the need for vastly better detection, even for large objects that are more easily detected. Note that the orbital phase of a comet like C2020/F3 and the orbital phase of the Earth, were there be an impact would be critical. Analyzing the orbit of C2020/F3 shows that is has heliocentric speeds at solar perihelion of about 78 km/s and heliocentric speed when closest to the Earth of about 53 km/s, yet when it crosses closest to the Earth’s orbit (not the Earth) it has a heliocentric speed of about 70 km/s, but in a direction such that if the Earth were in a phase in its orbit around the Sun to collide with C2020/F3 the Earth closing speed would be close to 100 km/s, or a devastating speed of impact.

Predicting the future of cosmic threats is easy for known threats, but not so easy for unknown threats. This is an obvious statement, but comet NEOWISE is a very recent and a very good example of what we lack, namely good situational awareness. Sungrazer comets, like NEOWISE, are particularly difficult to detect and serve as a warning.

\subsection{Comparison to Halley's Comet}

One of the most famous comets is Halley’s comet, discovered in 1758, which comes by the Earth about every 75 years. The next pass will be in July 2061, with a closest approach of about 0.064 AU (25 LD). It has a nucleus size of about $15\times8$ km, with a mean diameter of about 11 km and an estimated density of 0.6 g/cm$^3$, with a total mass of about $2.2\times10^{14}$ kg. Note that this density is close to the density we assumed (0.65 g/cm$^3$) for our comet threat. This lowers the gravitation binding energy ($\sim\rho^2$) as well as the dispersal energy ($\sim\rho$). We know relatively little about the detailed macroscopic structure of comet nuclei, but from our limited knowledge the estimated mean density is less than 1 g/cm$^3$, with an average density of $0.6\pm0.2$ g/cm$^3$ \cite{weissman_structure_2008}. They appear to be rubble piles in composition. The rubble pile nature and low mean density should make disassembly easier than more cohesive structures found in some asteroids, though larger asteroids are also likely rubble piles in nature, in general. Halley’s comet is another example of a possible comet ``threat'' that could be mitigated.

\subsection{Large Scale Porosity of Comet Nuclei and Coupling Coefficient}

One of the complexities of computing the coupling coefficient between an NED and a comet nucleus or a large asteroid is the large scale porosity of the material. The issue is the NED is most efficient at coupling the initial NED blast into the desired subsequent low speed dispersal, if the NED is ``well-tamped.'' This means the initial high temperature explosion which vaporizes the local surrounding material will build up a large volume of high pressure gas from the vaporized material. This gas, like that of an air bag in a car, will ``push'' the overlaying material outward and form the coupling between the NED explosive energy and the mechanical energy of the comet nucleus/asteroid. If there are large pores in the nucleus structure, this will lead to poor gas coupling and hence overall poorer coupling efficiency. For example, the coupling efficiency in an NED air blast on the Earth is about 50\% as there are essentially no pores in the air. The detailed effects are much more complex due to various energy and heat transfer mechanisms. This is an area which requires detailed simulations for a wide variety of cases. To put this in perspective, the use of ``wicker basket'' designs for conventional bomb disposal systems is precisely designed to minimize the coupling coefficient between the explosive and the mechanical structure holding the explosive inside it by creating large porosity which lets out the gas formed from the explosive and the air driven shock wave quickly. 

\subsection{Multiple Comet Nuclei}

The structure of comet nuclei is largely unknown and is very likely highly variable. There may be multiple loosely bound nuclei which could be advantageous in a fragmentation strategy, or it could be detrimental. Having multiple interceptors temporally spaced apart by hours could be extremely useful for independent targeting of residual large fragments still needing to be fragmented.

\subsection{Relation to Science Fiction Movies Using the Space Shuttle}

In some SciFi movies involving planetary defense, the Space Shuttle system has been shown as a vehicle to reach a threat to mitigate it. The Shuttle has large negative characteristic energy $C_3$ and cannot escape the Earth’s gravity and can only reach LEO. Interception at LEO for any sizeable existential threat is not viable, even if it used all of the Earth’s nuclear arsenal. While it makes for good cinema, it is not a feasible approach. An intercept at LEO altitudes of (say) 600 km altitude would be 15 seconds prior to impact for a threat closing speed of 40 km/s, requiring a transverse fragment speed of $\sim1000$ km/s for the fragments to clear the Earth. From Figure \ref{fig:disassemblyenergy}, it is clear that such speeds for a 10 km diameter, density 0.65 g/cm$^3$ comet would require about 40 billion Mt (40 Pt) of energy at 100\% coupling efficiency, vastly more ($\sim6$ million times) than the entire Earth’s arsenal. Short term (LEO interdiction) of existential bolide threats is simply not an option.

\subsection{An Asteroid the Size of Texas}

Let’s suppose that there was an asteroid the size of Texas that was going to impact Earth in less than a month. What could we do? First of all, how large is Texas? Answer: approximately 830 km diameter gives the proper projected area. Are there any asteroids this size? Answer: Yes, Ceres, the first asteroid ever detected, is about 930 km in diameter. It was discovered on January 1, 1801 and was visited by the Dawn mission in 2015. OK, so how much energy does it take to just gravitationally de-bind it assuming a typical rocky asteroid density of 2.6 g/cm$^3$? Answer:  About 14 Pt TNT, or 14 million Gigatons TNT, or about 2 million times more energy that all the Earth’s nuclear arsenal. What do you do now? You are going to need some die hard to get you out of this one. A couple of options: a) party, b) move to Mars or the Moon to party, c) do what they did in Chicken Run during take-off.

\subsection{Don't Forget to Look Down}

In the KT extinction, approximately 80\% of all animals went extinct. Some subterranean and ocean-based species did survive. Given the high heat capacity of water and of soil, it is not unreasonable to assume that a strategy of taking life underwater or underground would be a wise ``civil defense-based'' strategy to assure some survival of the human and other species. Multiple year food and water reserves would be required, but it is not unreasonable to imagine if primary defense fails, much as bomb shelters were and still are used. This paper is not focused on this coping mechanism, but it is an obvious layer of defense, as is in-orbit and off-planet life and seed banks. Being a multi-planet species as has been discussed in recent decades is another long term survival strategy in a layered defense system that includes active approaches to survival.

\section{Conclusion}

We have shown that for the extreme case of a 6 month warning of the impact of a 10 km diameter comet or asteroid, humanity could, in theory, defend itself with an array of nuclear (NED) penetrators launched 5 months prior to impact and an intercept one month prior to impact with a 5m/s fragmentation dispersal speed (at $\infty$). Using the same methodology we have outlined in our recent terminal planetary defense (PI) paper (our threat mitigation technique which works via hypervelocity penetrator array fragmentation and dispersal, but “upgraded” to use NED’s), humanity could prevent going the way of the dinosaurs who never took a physics class and failed to fund planetary defense. We note that the assumption of six-month notice is, in general, highly unlikely, especially for 10 km diameter asteroids,  given our ability to track and predict the orbital parameters of large diameter targets, though the case of comet NEOWISE, discovered in 2020 with only a four-month warning is a cautionary tale to be considered. The purpose of this paper is to show that even in relatively extreme short-term warning cases we can still effectively respond and mitigate the threat \textit{if} we prepare ahead. Though the numbers may seem daunting, it is not outside the realm of possibility even at this point in human technological development. This gives us hope that a robust planetary defense system is possible for even short notice existential threats such as we have outlined. Ideally, we would never be in this situation, but better ready than dead.

\section*{Acknowledgments}
We gratefully acknowledge funding from the NASA California Space Grant NASA NNX10AT93H and funding from the Emmett and Gladys W Technology Fund. 

\section*{Supplementary Material}

This paper and additional papers, simulations, and visualizations can be found on our website:
\texttt{https://www.deepspace.ucsb.edu/projects/pi-terminal-planetary-defense}. This area will be updated as new material and interactive simulations are developed. 

%Bibliography
\bibliographystyle{unsrt}  
\bibliography{Dont-Forget-to-Look-Up-references}  

\begin{thebibliography}{10}

\bibitem{popova_chelyabinsk_2013}
Olga~P. Popova et~al.
\newblock Chelyabinsk {Airburst}, {Damage} {Assessment}, {Meteorite}
  {Recovery}, and {Characterization}.
\newblock {\em Science}, 342(6162):1069--1073, November 2013.
\newblock Publisher: American Association for the Advancement of Science
  Section: Research Article.

\bibitem{glasstone_effects_1977}
S.~Glasstone and P.~J. Dolan.
\newblock The {Effects} of {Nuclear} {Weapons}. {Third} edition.
\newblock Technical Report TID-28061, Department of Defense, Washington, D.C.
  (USA); Department of Energy, Washington, D.C. (USA), January 1977.

\bibitem{brown_flux_2002}
P.~Brown, R.~E. Spalding, D.~O. ReVelle, E.~Tagliaferri, and S.~P. Worden.
\newblock The flux of small near-{Earth} objects colliding with the {Earth}.
\newblock {\em Nature}, 420(6913):294--296, November 2002.
\newblock Bandiera\_abtest: a Cg\_type: Nature Research Journals Number: 6913
  Primary\_atype: Research Publisher: Nature Publishing Group.

\bibitem{alvarez_extraterrestrial_1980}
Luis~W. Alvarez, Walter Alvarez, Frank Asaro, and Helen~V. Michel.
\newblock Extraterrestrial {Cause} for the {Cretaceous}-{Tertiary}
  {Extinction}.
\newblock {\em Science}, 208(4448):1095--1108, June 1980.
\newblock Publisher: American Association for the Advancement of Science.

\bibitem{taylor_formation_1950-1}
Geoffrey~Ingram Taylor.
\newblock The formation of a blast wave by a very intense explosion. - {II}.
  {The} atomic explosion of 1945.
\newblock {\em Proceedings of the Royal Society of London. Series A.
  Mathematical and Physical Sciences}, 201(1065):175--186, March 1950.
\newblock Publisher: Royal Society.

\bibitem{taylor_formation_1950}
Geoffrey~Ingram Taylor.
\newblock The formation of a blast wave by a very intense explosion {I}.
  {Theoretical} discussion.
\newblock {\em Proceedings of the Royal Society of London. Series A.
  Mathematical and Physical Sciences}, 201(1065):159--174, March 1950.
\newblock Publisher: Royal Society.

\bibitem{zeldovich_theory_1950}
Y.~Zeldovich.
\newblock On the theory of the propagation of detonation in gaseous systems.
\newblock {\em undefined}, 1950.

\bibitem{lubin_asteroid_2022}
Philip Lubin and Alexander~N. Cohen.
\newblock Asteroid {Interception} and {Disassembly} for {Terminal} {Planetary}
  {Defense}.
\newblock {\em Advances in Space Research}, 2022.

\bibitem{lubin_pi_2022}
Philip Lubin.
\newblock {PI} -- {Terminal} {Planetary} {Defense}.
\newblock {\em arXiv:2110.07559 [astro-ph, physics:physics]}, January 2022.
\newblock arXiv: 2110.07559.

\bibitem{lubin_planetary_2021}
Philip Lubin and Alexander~N. Cohen.
\newblock Planetary {Defense} {Is} {Good} - but {Is} {Planetary} {Offense}
  {Better}?
\newblock {\em Scientific American}, October 2021.

\bibitem{weissman_structure_2008}
Paul~R. Weissman and Stephen~C. Lowry.
\newblock Structure and density of cometary nuclei.
\newblock {\em Meteoritics \& Planetary Science}, 43(6):1033--1047, 2008.
\newblock \_eprint:
  https://onlinelibrary.wiley.com/doi/pdf/10.1111/j.1945-5100.2008.tb00691.x.

\bibitem{national_research_council_effects_2005}
{National Research Council}.
\newblock {\em Effects of {Nuclear} {Earth}-{Penetrator} and {Other}
  {Weapons}}.
\newblock The National Academies Press, Washington, DC, 2005.

\end{thebibliography}
%\nocite*

\end{document}